\documentclass[conference]{IEEEtran}
\IEEEoverridecommandlockouts

\usepackage{cite}
\usepackage{amsmath,amssymb,amsfonts}
\usepackage{algorithmic}
\usepackage{graphicx}
\usepackage{textcomp}
\usepackage{url}
\usepackage{graphicx}
\usepackage{amssymb,amsmath,bm}
\usepackage{textcomp}
\usepackage{booktabs}
\usepackage{hyperref}
\usepackage{verbatim}
\usepackage{multirow}
\usepackage[table,xcdraw]{xcolor}
\usepackage{array}
\usepackage{tabularx}
\usepackage{bbding}
\usepackage{xspace}
\usepackage{color} 
\newcolumntype{Y}{>{\centering\arraybackslash}X}
\newcommand{\norm}[1]{\left\lVert#1\right\rVert}
\def\BibTeX{{\rm B\kern-.05em{\sc i\kern-.025em b}\kern-.08em
    T\kern-.1667em\lower.7ex\hbox{E}\kern-.125emX}}
\begin{document}

\title{ComplexDec: A Domain-robust High-fidelity Neural Audio Codec with Complex Spectrum Modeling\\
}

\author{\IEEEauthorblockN{\begin{tabular}{c}{Yi-Chiao Wu, Dejan Marković, Steven Krenn, Israel D. Gebru, and Alexander Richard}\end{tabular} }
\IEEEauthorblockA{\textit{Codec Avatars Lab}, \textit{Meta}, Pittsburgh PA, USA }
}

\maketitle

\begin{abstract}
Neural audio codecs have been widely adopted in audio-generative tasks because their compact and discrete representations are suitable for both large-language-model-style and regression-based generative models. However, most neural codecs struggle to model out-of-domain audio, resulting in error propagations to downstream generative tasks. In this paper, we first argue that information loss from codec compression degrades out-of-domain robustness. Then, we propose full-band 48~kHz ComplexDec with complex spectral input and output to ease the information loss while adopting the same 24~kbps bitrate as the baseline AuidoDec and ScoreDec. Objective and subjective evaluations demonstrate the out-of-domain robustness of ComplexDec trained using only the 30-hour VCTK corpus.
\end{abstract}


\section{Introduction}
Conventional DSP-based audio codecs~\cite{codec1976, codec1986, codec1994-1, codec1994-2} focus on aggressive compressions of audio signals to efficiently store and transmit audio data but markedly sacrifice audio quality. Modern DSP-based audio codecs greatly improve audio quality with a lighter compression ratio from $2\times$ (lossless codecs~\cite{mpeg4, flac}) to $10\times$ (lossy codecs~\cite{ opus, amrwb, evs}) thanks to advanced codec technologies. Recently, many neural codecs~\cite{soundstream, encodec, audiodec} adopting an autoencoder (AE) with residual-vector-quantizer (RVQ)~\cite{rvq} architecture have been proposed to provide discrete audio tokens for large-language-model (LLM)-based audio generations~\cite{gslm, audiolm, valle, musicgen}. However, while DSP-based codecs usually work well for arbitrary audio,  data-driven neural codecs are vulnerable to unseen audio, which usually results in serious error propagations to the downstream audio generations. Scaling up the training data is helpful, but large-scale training is difficult and resource-consuming, and domain balance to avoid model bias is also challenging.

To investigate out-of-domain robustness, we revisit the network architecture and argue that information loss caused by temporal and dimensional compressions degrades robustness. Specifically, an encoder usually projects a high-temporal-resolution waveform into a low-temporal-resolution high-dimensional space and then also compresses the embedding dimension for the codebook learning stability. To ease the temporal and dimensional compressions, we propose ComplexDec to code speech in the complex spectral domain.  Because of the low temporal resolution (e.g. 150~Hz) of complex spectra, ComplexDec adopts a fully convolutional architecture without downsampling and upsampling layers to bypass the temporal compression. Because of the low spectral dimension setting (e.g. 256-dim), dimensionality reductions are unnecessary. Extracting complex spectra by short-time Fourier transform (STFT) also causes very limited information loss. Compared to waveform-based baselines~\cite{audiodec, scoredec}, ComplexDec eases information loss to improve out-of-domain robustness by avoiding most temporal and dimensional compressions. 

Objective and subjective evaluations are conducted using the reading-style VCTK~\cite{vctk2017}  corpus as the training set and the expressive EARS~\cite{ears} corpus as the testing set to demonstrate the out-of-domain robustness of ComplexDec. To the best of our knowledge, ComplexDec is the first paper to explore the out-of-domain robustness of neural codecs, although several codecs~\cite{apcodec, funcodec} also work in the complex spectral domain.

\section{Method}
\subsection{Problem Formulation}
Given a standard waveform-domain RVQAE codec~\cite{soundstream, encodec, audiodec } with a 150~Hz frame rate and 16 10-bit codebooks operating on 24~kbps ($150 \times 16 \times 10$) for 48~kHz speech coding, the \textbf{compression ratio} is $\frac{48000}{150 \times H}$, where $H$ is the dimension of the codes. Since the bitrate is only related to the temporal resolution (frame rate) of the codes when given a fixed codebook size, we can keep the same bitrate and reduce the compression ratio to ease the information loss by using a high $H$. However, to avoid unstable codebook learning, standard neural codecs such as AudioDec still compress the code dimension, which results in marked information loss. Descript-audio-codec (DAC)~\cite{dac} adopts a handcraft code factorization in RVQ to ease the information loss by adopting high-dimensional code embeddings (e.g. 1024-dim) while maintaining the codebook learning stability by performing a low-dimensional code lookup. However, in addition to the extra engineering efforts and training costs, the high dimensional representation is not preferable to regression-based audio generations because of the markedly increased audio modeling difficulties and memory requirements.

\subsection{Model Overview}
To ease the information loss and avoid the high dimension issue, ComplexDec codes audio in the complex spectral domain with a 320 hop length, 510 STFT size, and Hann window, which results in a 150~Hz frame rate and 256 embedding dimension. With a 24~kbps bitrate, ComplexDec also adopts an RVQAE architecture with 16 10-bit codebooks (8 for real and 8 for imaginary) but without downsampling and upsampling layers. Compared to the waveform-based baseline AudioDec using a 64-dim code, the 1.25 \textbf{compression ratio} of ComplexDec is much lower than the 5 \textbf{compression ratio} of AudioDec. Compared to 24~kbps DAC with a low 0.3125 \textbf{compression ratio} and high 1024-dim codes, ComplexDec adopts more preferable 256-dim codes. 

\begin{figure}[t]
  \centering
  \includegraphics[width=0.75\linewidth]{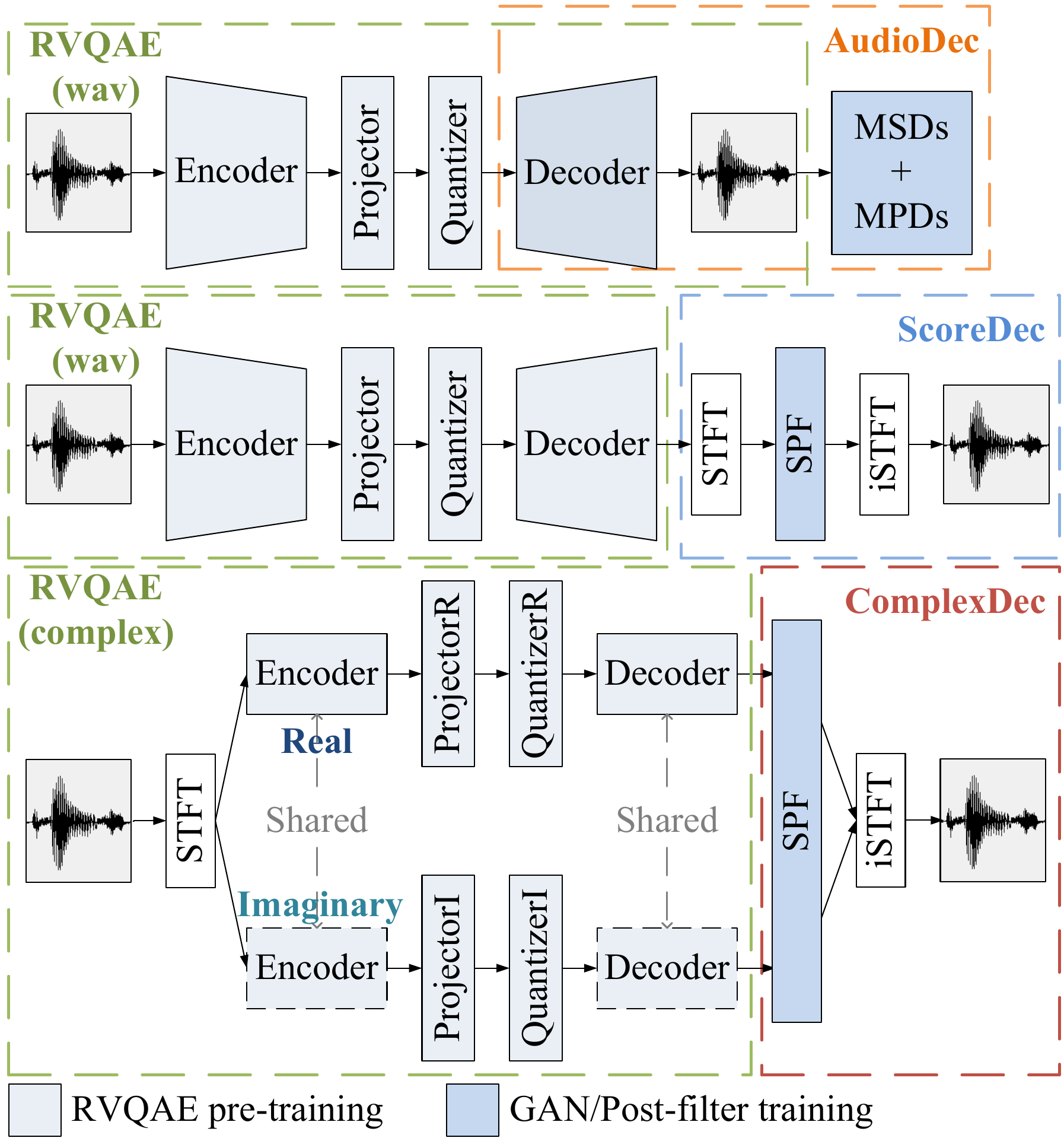}
  \caption{Model architecture comparison.}
  \label{fig:complexdec}
\end{figure}

\subsection{Model Architecture}
As shown in Fig.~\ref{fig:complexdec}, ComplexDec consists of two RVQAEs for real and imaginary spectra. The encoder and decoder are shared but the codebooks are independent, which is essential for the reconstruction quality. A score-based post-filter (SPF)~\cite{scoredec} is adopted to refine the decoded complex spectra. 

The encoder consists of a 1-dimensional convolutional layer (Conv1D) with kernel size 7, four encoder blocks, and an additional Conv1D with kernel size 3. The channel number of all Conv1Ds is 256. Each encoder block consists of three \textit{residual units} following a Conv1D with kernel size 2. Each \textit{residual unit} comprises two Exponential Linear Unit (ELU)-dilated Conv1D combinations with a residual connection. The dilations are [1, 3, 9] and the kernel size is 7. The decoder has a mirrored architecture but replaces the dilated Conv1Ds with transpose Conv1Ds. The RVQAEs are trained using spectral ($L_{MSE}$ and $L_{MAE}$), multi-resolution mel-spectral ($L_{Mel}$), and commitment VQ ($L_{VQ}$) losses. The codebooks are updated using the exponential moving averages (EMA)~\cite{vqvae} of the encoded/residual representations. Specifically, given the input spectrum $\boldsymbol{x}=\boldsymbol{x_r}+i\boldsymbol{x_i}$ and reconstructed spectrum $\boldsymbol{\hat{x}}=\boldsymbol{\hat{x}_r}+i\boldsymbol{\hat{x}_i}$, the $L_{MSE}$ loss is formulated as
\begin{align}
L_{MSE}=
\mathbb{E}\left[
\frac{\norm{\boldsymbol{x_r}-\boldsymbol{\hat{x}_r}}_{2}
+\norm{\boldsymbol{x_i}-\boldsymbol{\hat{x}_i}}_{2}}{2}
\right],
\label{eq:lmse}
\end{align}
and the $L_{MAE}$ loss is formulated as
\begin{align}
L_{MAE}=\mathbb{E}\left[\norm{\boldsymbol{x}-\boldsymbol{\hat{x}}}_{1}\right].
\label{eq:lmae}
\end{align}
An additional inverse STFT module is adopted for calculating the $L_{mel}$ with 80-dim mel-spectra extracted using [50, 120, 240] hop lengths and [512, 1024, 2048] STFT sizes.

In addition, the SPF is trained using score-matching~\cite{score_match} following ScoreDec~\cite{scoredec, sgmse1, sgmse2}. Specifically, given a natural and coded complex spectral pair $(\boldsymbol{x}_0, \boldsymbol{\hat{x}})$, the Ornstein-Uhlenbeck variance exploding (OUVE)~\cite{ouve} stochastic forward process of the SPF is defined as
\begin{align}
\mathrm{d}\boldsymbol{x}_t=
\underbrace{\gamma(\boldsymbol{\hat{x}}-\boldsymbol{x}_t)}_{:=f(\boldsymbol{x}_t,\boldsymbol{\hat{x}})}{\mathrm{d}}t
+
\underbrace{\begin{bmatrix}\sigma_{\text{min}}(\frac{\sigma_{\text{max}}}{\sigma_{\text{min}}})^t\sqrt{2\log(\frac{\sigma_{\text{max}}}{\sigma_{\text{min}}})}\end{bmatrix}}_{:=g(t)}\mathrm{d}\textbf{w},
\label{eq:sde}
\end{align} 
where $t\in[0, T]$ is a diffusion time step, $\textbf{w}$ is a standard Wiener process, $f(\boldsymbol{x}_t,\boldsymbol{\hat{x}})$ is the drift function, $g(t)$ is the diffusion coefficient, and $\gamma$ and $(\sigma_{\text{min}}, \sigma_{\text{max}})$ are the constant hyperparameters. Given the score function $\nabla_{\boldsymbol{x}_t}\log p_t(\boldsymbol{x}_t)$ as $\boldsymbol{s}$ and the time-reversed Wiener process ${\Bar{\textbf{w}}}$, the corresponding reverse stochastic differential equation (SDE)~\cite{rdem, sgm} is
\begin{align}
\mathrm{d}\boldsymbol{x}_t=
\begin{bmatrix}-f(\boldsymbol{x}_t,\boldsymbol{\hat{x}})+g(t)^2\boldsymbol{s}\end{bmatrix}{\mathrm{d}}t
+
g(t)\mathrm{d}{\Bar{\textbf{w}}}.
\label{eq:rsde}
\end{align}
Since $\boldsymbol{x_t}$ in the Gaussian process (Eq.~\ref{eq:sde}) follows a normal distribution with mean $\mu(\boldsymbol{x}_0, \boldsymbol{\hat{x}}, t)$ and variance $\sigma(t)^2$, $\boldsymbol{x_t}$ can be computed from a sampled Gaussian noise $\boldsymbol{z}$ by
\begin{align}
\boldsymbol{x}_t=
\mu(\boldsymbol{x}_0, \boldsymbol{\hat{x}}, t)+\sigma(t)\boldsymbol{z}.
\label{eq:xt}
\end{align}
The score function $\boldsymbol{s}$, which is formulated as
\begin{align}
\nabla_{\boldsymbol{x}_t}\log p_t(\boldsymbol{x}_t|\boldsymbol{x}_0,\boldsymbol{\hat{x}})=
-\frac{\boldsymbol{x}_t-\mu(\boldsymbol{x}_0, \boldsymbol{\hat{x}}, t)}{\sigma(t)^2},
\label{eq:score}
\end{align}
can be estimated using a neural network $\boldsymbol{s}_\theta$ trained by the score-matching objective function
\begin{align}
\mathop{\arg\min}_{\theta}\mathbb{E}_{\boldsymbol{x}_t|(\boldsymbol{x}_0,\boldsymbol{\hat{x}}),\boldsymbol{\hat{x}},\boldsymbol{z},\boldsymbol{t}}
\left[\norm{\boldsymbol{s}_\theta(\boldsymbol{x}_t, \boldsymbol{\hat{x}}, t)+\frac{\boldsymbol{z}}{\sigma(t)}}_{2}^2\right].
\label{eq:score_match}
\end{align}
Given a well-trained $\boldsymbol{s}_\theta$, the SPF adopts the predictor-corrector samplers using \textit{reverse diffusion sampling} as the predictor and \textit{annealed Langevin dynamics} as the corrector~\cite{sgm} for the reverse SDE. The SNR parameter of the corrector is 0.5 and the number of the reverse steps is 30. Moreover, the spectra in the diffusion process are modulated and demodulated as
\begin{align}
\boldsymbol{x^\prime}=\beta|\boldsymbol{x}|^\alpha e^{i\angle(\boldsymbol{x})},
\label{eq:am}
\end{align}
and
\begin{align}
\boldsymbol{x}=\beta^{-1}|\boldsymbol{x^\prime}|^{\frac{1}{\alpha}} e^{i\angle(\boldsymbol{x^\prime})},
\label{eq:dm}
\end{align}
where $\angle(\cdot)$ denotes the angle of a complex number, $\alpha=0.5$ is an amplitude companding constant, and $\beta=0.15$ is a scaling constant to normalize the amplitudes roughly within $[0,1]$.

The SPF score model adopts a U-Net style Noise Conditional Score Network (NCSSN++) architecture~\cite{sgm} and takes the real and imaginary spectra as two channels. Given the 4-channel 256$\times$256 input of complex spectra $\boldsymbol{x}_0$ and $\boldsymbol{\hat{x}}$, the NCSSN++ first gradually projects the feature maps into spaces with lower resolutions and higher channel numbers (e.g. 256-channel 4$\times$4 in the middle) and then projects the feature maps back to 256$\times$256 in a mirrored manner. Skip connections and progressive conditional structures are also integrated. The diffusion time steps are incorporated into the network using the Fourier embeddings technique~\cite{atten}.

\section{Experiments}
\label{sec:experiment}

\subsection{Experimental Setting}
We found that most neural codecs are trained using reading-style speech and suffer significant quality degradation when coding expressive speech so this paper focuses on speech coding. The reading-style VCTK~\cite{vctk2017}-derived Valentini~\cite{noisyvctk} clean subset and the expressive EARS-part-1 clean subset~\cite{ears} were adopted. Both subsets are full-band 48~kHz English datasets. Valentini-clean includes 84 gender-balanced speakers for training, and each speaker has around 400 utterances. EARS-part-1 includes 25 speakers for training and two speakers (one male and one female) for testing, and each speaker attains around 25 mins of data. The total length of the training data of Valentini-clean and EARS-part-1 are respectively around 30 and 10 hours. EARS is an expressive dataset including 22 emotion and 7 speaking types. VCTK is recorded in a semi-anechoic chamber using consumer devices while EARS is recorded in an anechoic chamber with high-end microphones. The quite different speakers, speech types, environments, and channel effects make the out-of-domain speech coding challenging.

The training weights of $L_{VQ}$, $L_{Mel}$, $L_{MSE}$, and $L_{MAE}$ were set to 1.0, 45, 200, and 200 in this paper. The batch size was 16, the batch length was 96,000, and the learning rate was $10^{-4}$ without any decay for the RVQAE training with Adam optimizers~\cite{adam}. The 500~k-step training took an NVIDIA A100 GPU for 7 hours. AudioDec v1\footnote{\label{repo}\url{https://github.com/facebookresearch/AudioDec}}  and ScoreDec, which is a symmetric AudioDec-variant with an SPF, were selected as the baselines because of the similar RVQAE architecture. The AudioDec and ScoreDec RVQAEs followed similar training settings but without the complex $L_{MSE}$ and $L_{MAE}$ and utilized a single-resolution Mel loss. Both the ScoreDec and ComplexDec SPFs followed the SGMSE+~\cite{sgmse1, sgmse2} repository\footnote{\url{https://github.com/sp-uhh/sgmse}} but changed the sampling rate to 48~kHz. The 161-epoch SPF training took four A100 GPUs for 8 hours.

To demonstrate the domain robustness, both in-domain and out-of-domain evaluations were conducted. For the in-domain evaluations, codecs were trained using both the Valentini-clean and EARS-part-1 training sets and tested on the EARS-part-1 test set. For the out-of-domain evaluations, codecs were trained only on the Valentini-clean training set. Moreover, the open-source 24~kHz and 48~kHz Encodec~\cite{encodec} and 24~kHz DAC~\cite{dac}, which are trained on tons of diverse audio data, were also included to demonstrate that solely scaling up the training data still suffers the degradations caused by information loss. The bitrates of all neural codecs were set to 24~kbps. Since DSP-based codecs are not data-driven and usually robust to arbitrary audio, we evaluated only these neural codecs.

\begin{table}[t]
\caption{Objective evaluations of codecs w/ 24~kbps}
\label{tb:objective}
\fontsize{8pt}{9.6pt}
\selectfont
{%
\begin{tabularx}{1.0\columnwidth}{@{}p{0.75cm}p{1.05cm}<{\centering}YYp{0.9cm}<{\centering}p{0.9cm}<{\centering}@{}}
\toprule
& $f_s$ (kHz) & Wav($\times$10\textsuperscript{-3})$\downarrow$ & SI-SDR$\uparrow$ &STOI$\uparrow$ &PESQ$\uparrow$ \\ \midrule
&& \multicolumn{4}{c}{\cellcolor[HTML]{F2F2F2}In-domain} \\
AudioDec      & 48 &1.61   &-19.62 & 0.84   & 3.28   \\
ScoreDec      & 48 &0.36   &5.66   & \textbf{0.92}   & 3.63   \\
\textbf{Ours} & 48 &\textbf{0.05}   &\textbf{13.69}  & 0.90   & \textbf{3.70}   \\
\midrule
&& \multicolumn{4}{c}{\cellcolor[HTML]{F2F2F2}Out-of-domain} \\
AudioDec      & 48 &1.49   &-16.75 & 0.77   & 1.72   \\
ScoreDec      & 48 &0.59   &1.89   & 0.87   & 2.49   \\
\textbf{Ours} & 48 &\textbf{0.06}   &\textbf{10.74}  & \textbf{0.88}   & \textbf{3.46}   \\
\midrule
&& \multicolumn{4}{c}{\cellcolor[HTML]{F2F2F2}Open-source (out-of-domain)} \\
Encodec       & 48 &0.10   &11.07  & 0.85   & 2.98   \\
Encodec       & 24 &0.21   &7.01   & 0.78   & 2.49   \\
DAC          & 24 &\textbf{0.04}  &\textbf{12.99}  &\textbf{0.98}    &\textbf{4.36}   \\
\bottomrule
\end{tabularx}%
}
\end{table}

\subsection{Objective Evaluation}
All test data were first resampled to 24~kHz for calculating the waveform mean-square-error (Wav) and scale-invariant source-to-distortion ratio (SI-SDR)~\cite{sisdr} in dB to evaluate the reconstruction performance and then resampled to 16~kHz for speech intelligibility and quality evaluations using wide-band STOI~\cite{stoi} and PESQ~\cite{pesq}. As shown in Table~\ref{tb:objective}, ComplexDec achieves similar in-domain and out-of-domain performances while AudioDec and ScoreDec suffer serious degradation in the out-of-domain case. Although the Encodec models are trained using a huge amount of data, the 2.5 \textbf{compression ratio} with 128-dim codes still causes severe information loss and results in marked out-of-domain quality degradation. On the other hand, ComplexDec achieves similar reconstruction performance but worse STOI and PESQ compared to DAC, which has a $4\times$ latent dimensional and much more training data. The results demonstrate the high correlation between information loss and domain robustness. However, because of the lack of effective full-band (48~kHz) objective measurements, only the objective results of the downsampled speech are reported. To evaluate the full-band perceptual qualities, mean opinion score (MOS) evaluations were also conducted.

\begin{table*}[t]
\caption{Mean opinion scores w/ 95\% confidence internals (CI) of perceptual speech quality of codecs w/ 24kbps}
\label{tb:mos}
\fontsize{8pt}{9.6pt}
\selectfont
{%
\begin{tabularx}{2.0\columnwidth}{@{}p{1.1cm}YYYYYYYYYYYY@{}}
\toprule
& \multicolumn{2}{c}{\cellcolor[HTML]{F2F2F2}Natural} & \multicolumn{3}{c}{In-domain} & \multicolumn{3}{c}{Out-of-domain} & \multicolumn{3}{c}{Open-source (out-of-domain)}\\ \cmidrule(lr){2-3} \cmidrule(lr){4-6} \cmidrule(lr){7-9} \cmidrule(lr){10-12}
System & - & - & AudioDec & ScoreDec & \textbf{Ours} & AudioDec & ScoreDec & \textbf{Ours} & Encodec & Encodec & DAC \\ 
$f_s$ (kHz) & 48 & 24 & 48 & 48 & 48 & 48 & 48 & 48 & 48 & 24 & 24 \\ 
\midrule
Reading    & 4.33 & 3.88 & 3.33 & \textbf{4.50} & 4.08 & 1.42 & 2.25 & 4.08 & 2.58 & 2.96 & \textbf{4.17} \\ 
Expressive & 4.14 & 3.72 & 3.00 & \textbf{4.19} & 3.88 & 1.24 & 2.36 & \textbf{3.99} & 2.25 & 2.64 & 3.60 \\
Loud       & 4.33 & 3.75 & 2.83 & \textbf{4.58} & 4.42 & 1.08 & 2.08 & \textbf{4.25} & 2.08 & 2.25 & 3.83 \\
Whisper    & 4.00 & 4.17 & 3.08 & \textbf{4.08} & 3.50 & 3.58 & \textbf{3.83} & 3.58 & 3.17 & 2.50 & 3.25 \\ 
\midrule
Overall & 4.18$\pm$.15 & 3.80$\pm$.16 & 3.06$\pm$.15 & \textbf{4.28}$\pm$.12 & 3.93$\pm$.16
& 1.49$\pm$.16 & 2.46$\pm$.16 & \textbf{3.99}$\pm$.14 & 2.39$\pm$.17 & 2.65$\pm$.14 & 3.70$\pm$.16 \\
\bottomrule
\end{tabularx}%
}
\end{table*}

\begin{figure*}[t]
\fontsize{9pt}{9pt}
\selectfont
{%
\begin{tabularx}{2.0\columnwidth}{@{}p{0.1cm}XXXX@{}}
\rotatebox[origin=c]{90}{Frequency (Hz)}
& \includegraphics[width=0.5\columnwidth]{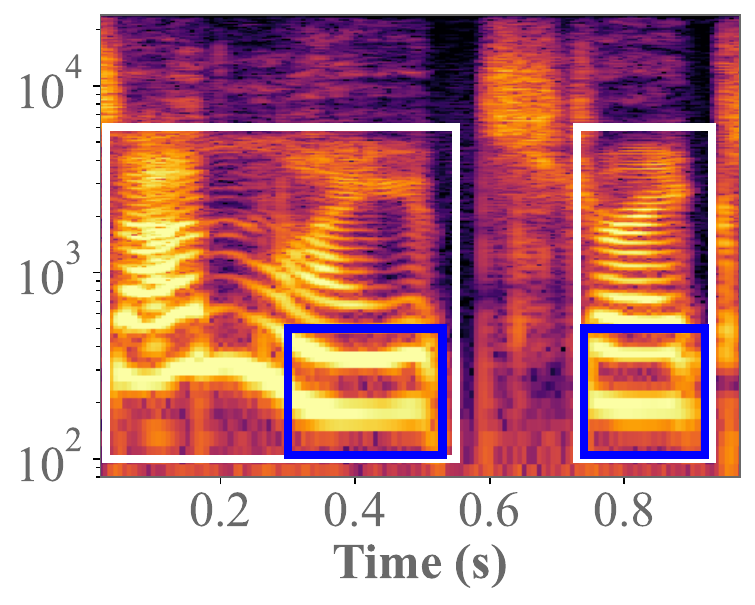}
& \includegraphics[width=0.5\columnwidth]{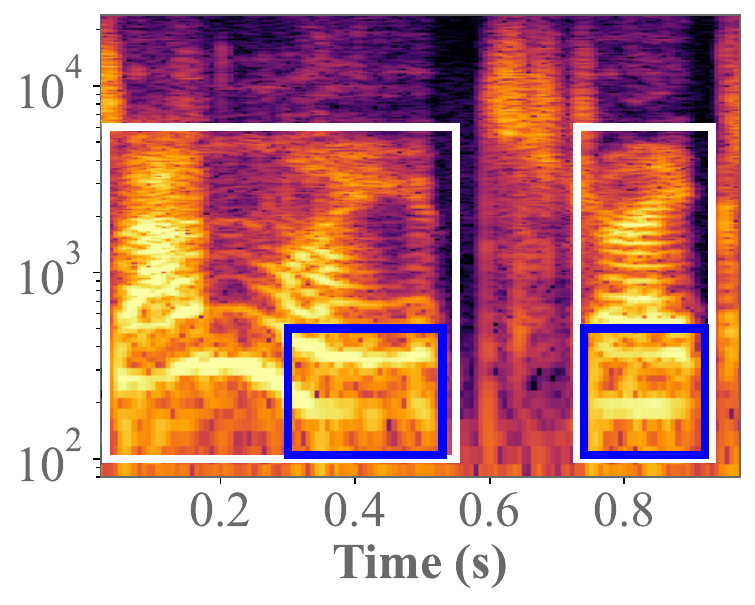}
& \includegraphics[width=0.5\columnwidth]{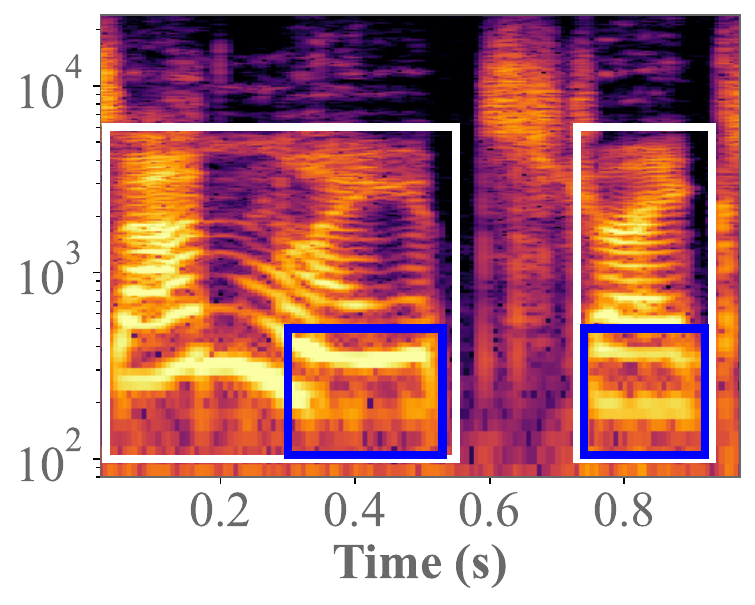}
& \includegraphics[width=0.5\columnwidth]{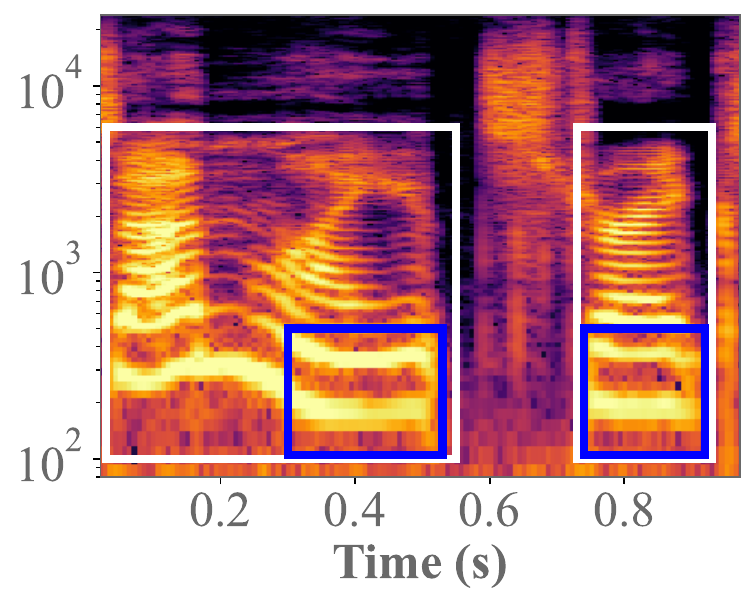}
\\

&  \centering\arraybackslash{(a) Natural}
&  \centering\arraybackslash{(b) AudioDec}
&  \centering\arraybackslash{(c) ScoreDec}
&  \centering\arraybackslash{(D) \textbf{ComplexDec}}
\\
\end{tabularx}%
}
\caption{Comparison of natural/reconstructed magnitude spectra of out-of-domain loud speech.}
\label{fig:mag}
\end{figure*}

\subsection{Subjective Evaluation}
As shown in Table~\ref{tb:mos}, ground truth 48~kHz and 24~kHz speech were also included in addition to the codec results. Two reading, six expressive, one loud, and one whisper utterances were selected for the MOS evaluations. Each utterance is around 8--20s, so the overall test data from all systems is around 25~mins. Our internal annotation team, which consists of 12 native or expert English speakers, took the subjective test in a quiet environment with headphones. The score range is 1 (very unnatural) -- 5 (very natural). The average scores with 95\% confidence intervals are reported.

Again, ComplexDec achieves similar in-domain and out-of-domain coding qualities while AudioDec and ScoreDec suffer significant degradation in coding the out-of-domain speech. ComplexDec also significantly outperforms the open-source Encodec models. The results demonstrate the out-of-domain robustness of ComplexDec, and the serious information loss cannot be fully compensated by the SPF or by solely increasing the training data. On the other hand, DAC also achieves impressive out-of-domain robustness because of its low compression ratio. However, the marked quality gap between ComplexDec and DAC shows that the perceptual quality difference between 48~kHz and 24~kHz speech cannot be reflected in the downsampled objective measurements. More details can be found on our demo page\footnote{\url{https://bigpon.github.io/ComplexDec_demo/}}.   

\subsection{Discussion}
To explore the out-of-domain speech coding difficulties, we plot the magnitude spectra in Fig~\ref{fig:mag}. The results show that AudioDec fails to reconstruct the harmonic structures and the blur spectrum results in hoarse speech. Considering the human voice source-filter model, the result indicates that AudioDec suffers low-quality voiced source signal modeling while maintaining only the waveform envelope.  The ScoreDec result also shows that although the SPF can slightly recover the blurry spectrum because of the diffusion nature, the missing harmonics cannot be well recovered. The same tendency can be observed in the subjective results that AudioDec and ScoreDec show surprise out-of-domain robustness in whisper coding because whisper is unvoiced speech without harmonic structures. The results also imply that the serious information loss makes the decoder vulnerable to modeling unseen long-term dependency, which crosses several sequential codes.

In addition, ComplexDec well preserves the harmonic structures below 6~kHz, and the corresponding perceptual qualities of the in-domain and out-of-domain reading, expressive, and loud speech coding are also similar. However, ComplexDec lost more high-frequency details than ScoreDec does. Compared to the 16 codebooks of ScoreDec for the quantization, the 8 codebooks of the real or imaginary spectrum of ComplexDec cause higher quantization errors, which result in a slightly greater loss of high-frequency details. However, given the much better domain robustness, ComplexDec still achieves much higher perceptual quality in out-of-domain coding.

On the other hand, ComplexDec achieves a 1.644 real-time factor (RTF) on an NVIDIA A100 80GB SXM GPU while the RTF of ScoreDec is 1.707. Compared to the 0.022 RTF of AudioDec, the much slower iterative SPF inference hinders ComplexDex from streaming applications, and we leave the fast inference for future work.

\section{Conclusion}
In this paper, we explore the domain robustness of neural codecs, which is the main weakness of current neural codecs compared to DSP-based codecs. After revisiting the neural codec architecture,  we propose that the information loss caused by the temporal and dimensional compressions degrade the out-of-domain robustness. Therefore, we propose ComplexDec to bypass these compressions by adopting the complex spectral input and output. The experimental results show significant out-of-domain robustness of ComplexDec in expressive speech coding even though the model was trained using only 30~hours of reading style speech. Compared to the strong open-source baselines trained using thousands of hours of data, the significantly better out-of-domain robustness of ComplexDec indicates a more promising direction than merely increasing the amount of training data. Furthermore, the low embedding dimensionality of ComplexDec is also preferable for the regression-based audio generations. For future work, easing the quantization errors, exploring streamable inference, and integrating ComplexDec into audio generative models are interesting directions. 

\bibliographystyle{IEEEbib}
\bibliography{mybib}

\end{document}